\documentclass[preprint, superscriptaddress,amsmath, nofootinbib]{revtex4-1}
\usepackage{graphicx}
\usepackage{latexsym}
\usepackage{slashed}
\usepackage{bm}
\usepackage{color}
\usepackage[colorlinks,citecolor=blue,urlcolor=blue,linkcolor=red]{hyperref}
\usepackage{diagbox}
\usepackage{makecell}
\pdfoutput=1
\usepackage{dcolumn}
\usepackage{bm}
\usepackage{multirow}
\usepackage{booktabs}
\usepackage{tabularx}
\begin{document}

\title{NANOGrav Signal and PBH from the Modified Higgs Inflation}
\def\slash#1{#1\!\!\!/}

\author{Kingman Cheung}
\email{cheung@phys.nthu.edu.tw}
\affiliation{Department of Physics, National Tsing Hua University, Hsinchu 30013, Taiwan} 
\affiliation{Center for Theory and Computation,
National Tsing Hua University, Hsinchu 30013, Taiwan}
\affiliation{Division of Quantum Phases and Devices, School of Physics, Konkuk University, Seoul 143-701, Republic of Korea}

\author{C.J. Ouseph}
\email{ouseph444@gmail.com}
\affiliation{Department of Physics, National Tsing Hua University, Hsinchu 30013, Taiwan} 
\affiliation{Center for Theory and Computation,
National Tsing Hua University, Hsinchu 30013, Taiwan} 

\author{Po-Yan Tseng}
\email{pytseng@phys.nthu.edu.tw}
\affiliation{Department of Physics, National Tsing Hua University, Hsinchu 30013, Taiwan} 
\affiliation{Center for Theory and Computation,
National Tsing Hua University, Hsinchu 30013, Taiwan}

\date{\today}

\begin{abstract}

This study investigates the classical Higgs inflation model with a modified Higgs potential featuring a dip. We examine the implications of this modification on the generation of curvature perturbations, stochastic gravitational wave production, and the potential formation of primordial black holes (PBHs). Unlike the classical model, the modified potential allows for enhanced power spectra and the existence of PBHs within a wide mass range $1.5\times10^{20}$ g -- $9.72\times10^{32}$ g. We identify parameter space regions that align with inflationary constraints and have the potential to contribute significantly to the observed dark matter content. Additionally, the study explores the consistency of the obtained parameter space with cosmological constraints and discusses the implications for explaining the observed excess in gravitational wave signals, particularly in the NANOGrav experiment. Overall, this investigation highlights the relevance of the modified Higgs potential in the classical Higgs inflation model, shedding light on the formation of PBHs, the nature of dark matter, and the connection to gravitational wave observations.
%


%
%
\end{abstract}

\maketitle

\section{Introduction}
Cosmological inflation is the most favorable theory of the early universe \cite{Guth:1980zm}. It not only explains the 
absence of a number of relics that should have existed from the Big Bang, but also provides the 
seeds for the growth of structures in the Universe. In the last two decades, people have been 
attempting to figure out the most promising candidate for cosmological inflation. 

Primordial black holes (PBHs) can also be created as a result of cosmological inflation, potentially generating 
PBHs from the seeds formed during the radiation- or matter-dominated epochs. Consequently, studying the 
formation and evolution of PBHs offers an effective means to investigate the early period of cosmology. 
The existence of primordial black holes was initially postulated by Zel'dovich and Novikov \cite{Zeldovich:1967lct}, and later supported by Hawking and Carr \cite{Hawking:1971ei,Carr:1974nx,Carr:1975qj}, suggesting that these hypothetical entities emerged in the early universe. 
According to the theory, PBHs may form in the regions with significant density perturbations. 
The primary motivation for studying PBHs lies in their potential 
as a natural candidate for dark matter. 
Despite recent observations imposing strict limitations on the abundance of PBHs, 
there exists a mass range, specifically from $10^{16}$ g to $10^{20}$ g, 
where PBHs could play a significant role in contributing to the overall dark matter content. 

Production of gravitational waves through the second-order effect is 
closely intertwined with the formation of PBHs, occurring simultaneously as 
certain modes re-enter the Hubble radius. These gravitational waves, once generated, 
propagate freely throughout subsequent epochs of the Universe due to their low interaction rate.

Millisecond pulsars (MPs) are featured by their stable rotating period which are comparable to the timing 
precision of atomic clocks. They are ideal astrophysical objects being utilized in the pulsar-timing arrays 
(PTAs) to probe the low-frequency gravitation waves (GWs) from nanohertz to microhertz. 
The NANOGrav collaboration has been observing 67 pulsars over 15 years and recently reported evidence 
for the correlations following Hellings-Downs pattern~\cite{NANOGrav:2023gor}, 
pointing to the stochastic GW as the origin. 
Furthermore, they confirmed the excess of the red common-spectrum signal with a strain 
amplitude of $\mathcal{O}(10^{-14})$ at the frequency $\simeq 3\times 10^{-8}$ Hz. 
To explain the GW signal, there are many plausible mechanisms and hypothetical candidates being proposed; 
in particular, the population of supermassive black-hole 
binarys~\cite{NANOGrav:2023gor,NANOGrav:2023hfp,Broadhurst:2023tus}, 
inflation scenarios~\cite{Niu:2023bsr,Antusch:2023zjk,Vagnozzi:2023lwo,You:2023rmn,Choudhury:2023kam}, 
cosmological first-order phase transition~\cite{DiBari:2023upq,Xiao:2023dbb,Salvio:2023ynn,Gouttenoire:2023bqy},
and alternative interpretations~\cite{Du:2023qvj,Wang:2023div,Babichev:2023pbf,Geller:2023shn}.

There have been numerous attempts to incorporate inflation into the standard model (SM) and theories beyond. 
The SM Higgs field has always been an intriguing candidate as the inflaton due to its lack of requirement for
additional scalar degrees of freedom. However, the minimal Higgs inflation model is not favored, and possibly
even ruled out, due to the fine-tuned value of the Higgs self-coupling constant, $\lambda$. 
To address this issue, a non-minimal coupling between the SM Higgs field and the Ricci scalar, 
$\mathcal{R}$, was introduced in an attempt to reduce the value of $\lambda$ \cite{Bezrukov:2007ep}. 
However, such attempts may potentially lead to violations of 
unitarity~\cite{Atkins:2010yg,Ren:2014sya,
Xianyu:2013rya}, although our current study does not focus on these infractions.

The fundamental Higgs inflation model \cite{Bezrukov:2007ep} fails to account for both the inflationary phase and the formation of PBHs
simultaneously. Numerous studies have explored both inflation and PBH formation 
within the framework of Higgs inflation by introducing new interactions to the Higgs field. By incorporating these additional interactions, it is possible to achieve a successful inflationary epoch while creating the conditions necessary for the generation of PHBs.
In this investigation, we examine a modified form of the Higgs potential that aims to address both the 
phenomenon of inflation and the formation of PBHs, and at the same time address the excess in GW signal reported by NANOGrav.

The paper is organized as follows. In Sec.~\ref{sec2}, we revisit the classical Higgs inflation model and 
demonstrate that it is incapable of generating the correct power spectrum required for 
PBH
formation. In Sec.~\ref{sec3}, we delve into possible modifications to the Higgs potential. Specifically, we
explore the introduction of a dip in the potential, which can accommodate PBH formation during the inflationary
scenario. We also discuss the viable parameter space by considering the characteristics of the dip for PBH
formation. Sections~\ref{sec4} and \ref{sec5} are dedicated to presenting the PBH abundance and gravitational
wave (GW) spectrum resulting from the modified Higgs potential. In Sec.~\ref{sec5}, we discuss the 
gravitational wave signals within this model and demonstrate that the obtained results can potentially 
explain the NANOGrav 15-year signal in a straightforward manner. 
We conclude with the implications of these findings.

\section{Revisiting Higgs Inflation Model}\label{sec2}
In this section, we revisit the classical Higgs Inflation model, which considers the SM Higgs Boson as a
promising candidate for inflation. The Higgs inflation model~\cite{Bezrukov:2007ep} was proposed 
a long time ago to bridge the gap between the two most successful models of physics: the standard model of
particle physics and the standard model of cosmology. 
Numerous studies~\cite{Maity:2016zeu, Rubio:2018ogq, He:2018gyf, Kamada:2012se, Ouseph:2020kcz, Lebedev:2011aq, Germani:2010gm} 
discussed the possibility of the SM Higgs as the inflaton in different contexts.

In our discussion, we focus on the simplest model of Higgs Inflation~\cite{Bezrukov:2007ep}. 
This model addresses inflation by introducing a non-minimal coupling, where the SM Higgs is coupled 
to the Ricci scalar $\mathcal{R}$ with a non-minimal coupling strength $\xi$. 
The effective action for this theory is given as
\begin{equation}\label{Eq.1}
\mathcal{S_J}=\int ~d^4x~\sqrt{-g}\big[-\frac{M_{\rm PL}^2+\xi h^2}{2}\mathcal{R} +\frac{\partial_{\mu}h\partial^{\mu}h}{2}-\frac{\lambda}{4}(h^2-v^2)^2 \big] \;.
\end{equation}
This Lagrangian has been studied in details in many works on inflation~\cite{Salopek:1988qh,Kaiser:1994vs,Komatsu:1999mt}. 
The scalar sector of the SM coupled to the gravity in a non-minimal way. Here the authors considered 
the unitary gauge $H=\frac{h}{\sqrt{2}}$ and neglected the interactions for the time being. 
An action in the Einstein frame was obtained by the conformal 
transformation~\cite{Kaiser:2010ps,Wald:1984rg} 
$\hat{g} _{\mu\nu}=\Omega g_{\mu\nu}$, where $\Omega=1+\frac{\xi h^2}{M^2_{\rm PL}}$. 
The conformal transformation can eliminate the non-minimal coupling to gravity.
This transformation leads to a non-minimal kinetic term for the
Higgs field. So, it is convenient to make the change to the new
scalar field $\phi$ with
\begin{equation}
 \label{Eq.2}
\frac{d\phi}{dh}=M_{\rm PL}\sqrt{\frac{\Omega}{\Omega^2}+\frac{3}{2}\frac{(\frac{d\Omega}{dh})^2}{\Omega^2}}
\end{equation}
The action in the Einstein frame is
\begin{equation}\label{Eq.3}
\mathcal{S_E}=\int d^4x
\sqrt{-\hat{g}}\big[-\frac{M^2_{\rm PL}}{2}\mathcal{\hat{R}}+\frac{\partial_{\mu}\phi\partial^{\mu}\phi}{2}-U(\phi)\big]. 
\end{equation}
The exponentially flat effective potential for the Higgs field is given by
\begin{equation}\label{Eq.4}
U(\phi)=\frac{\lambda M_{PL}^4 e^{-\frac{2 \sqrt{\frac{2}{3}} \phi }{M_{PL}}} \left(e^{\frac{\sqrt{\frac{2}{3}} \phi }{M_{PL}}}-1\right)^2}{\xi ^2}   
\end{equation}
The slow-roll parameters of this model of inflation can be expressed as the function of 
$h(\phi)$ as follows 
\begin{equation}\label{Eq.5}
 \epsilon=\frac{M^2_{\rm PL}}{2}\Bigg(\frac{\frac{\partial U}{\partial\phi}}{U}
  \Bigg)^2=\frac{4 M^4_{\rm PL}}{3\xi^2 h^4}
\end{equation}
\begin{equation}
\eta=M^2_{\rm PL}\Bigg(\frac{\frac{\partial^2U}{\partial\phi^2}}{U}\Bigg)=-\frac{4M^2_{\rm Pl}}{3\xi h^2}.
 \end{equation}
The slow roll ends when $\epsilon=1$, the field value at the end of inflation is given by,
\begin{equation}\label{Eq.7}
h_{end}=\Big(\frac{4}{3}\Big)^{\frac{1}{4}} \Big(\frac{M_{\rm PL}}{\sqrt{\xi}}\Big) .
\end{equation}
The number of $e$-folds that are required to change the field $h$ from $h_{int}$ to $h_{end}$ is given by 
\begin{equation}
 \label{Eq.8}
  N_{e}=\int_{h_{end}}^{h_{int}}\frac{1}{M_{\rm PL}^2}
  \Bigg(\frac{U}{\frac{\partial U}{\partial h}}
  (\frac{\partial\phi}{\partial h})^2\Bigg) dh \;.
\end{equation}
The field value $h_{int}$ at the beginning of inflation can be expressed as a function of 
$e$-folds 
\begin{equation}\label{Eq.9}
h_{int}=\frac{2M_{\rm PL}}{\sqrt{3\xi}}\big[\frac{\sqrt{3}}{2}+N_e \big]^{1/2}.
\end{equation}
The constraint over the Higgs self coupling constant $\lambda$ and the non-minimal coupling constant 
$\xi$ can be obtained from the 
$\rm COBE$ normalization $\frac{U}{\epsilon}=(0.027M_{\rm PL})^4$\cite{Lyth:1998xn}. 
Plugging Eq.~(\ref{Eq.9}) into the $\rm COBE$ normalization, 
we could express $\frac{\lambda}{\xi^2}$ as a function of the number of $e$-folds. 
For $N_e$=60, the constrain on  $\frac{\lambda}{\xi^2}=4.41026\times10^{-10}$. 
The inflationary parameters such as the scalar spectral index $n_s$  and the tensor-to-scalar ratio 
$r$ are defined as $n_s=1-6\epsilon+2\eta$, $r=16\epsilon$. With the number of e-folds  $N_e=60$  
this model gives the values of $r$(0.0032) and $n_s$(0.9633), which are well within the Planck bounds \cite{Planck:2015sxf} of $n_s=0.9677\pm0.0060$ and $r<0.11$ at 95$\%$ C.L.

The scalar power spectrum $\mathcal{P}_{\mathcal{R}}$ is defined as
\begin{equation}\label{Eq.10}
 \mathcal{P}_{\mathcal{R}}=\frac{1}{12\pi^2}\frac{U^3(\phi)}{M_{PL}^6 U'^2(\phi)} \;,
\end{equation}
where $U'(\phi)$ is the derivative of $U(\phi)$ with respect to $\phi$ and both $U'(\phi)$ and $U(\phi)$
are calculated at $\phi_{int}$\footnote{Eq.~~\ref{Eq.2} gives the relation connecting $\phi~\text{and}~h$, $\phi=\sqrt{\frac{3}{2}} \text{Mpl} \log \left(\frac{h^2 \xi }{\text{Mpl}^2}+1\right)$}. 
Recent CMB observations \cite{BICEP2:2018kqh} suggested the value of 
$\mathcal{P}_{\mathcal{R}}=2.1\times10^{-9}$ at the CMB pivot scale. Using 
Eq.~(\ref{Eq.10}) and Eq.~(\ref{Eq.9}) we can express the power spectra as a 
function of the number of e-folds with different choices of $\lambda/\xi^2$.
\begin{figure}[t!]
	\centering
	\includegraphics[width=12cm,height=8cm]{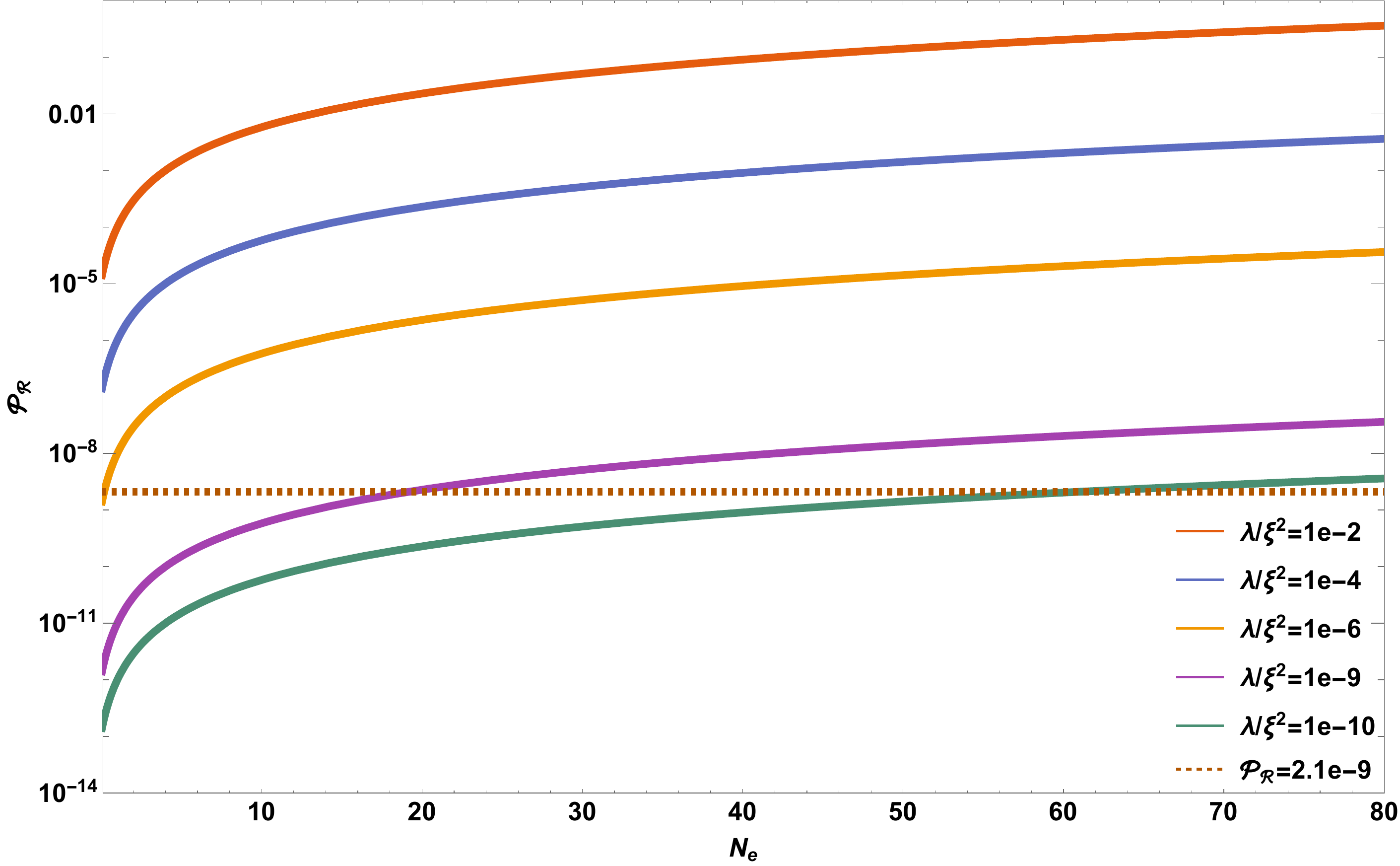}
	\caption{ \small \label{fig1}
	Scalar power spectra $\mathcal{P}_{\mathcal{R}}$  as the function of the number of e-folds $N_e$ with different choices of $\lambda/\xi^2$. }
\end{figure}
It is evident from Fig.~\ref{fig1} that $\mathcal{P}_{\mathcal{R}}$ 
attains a value of $2.1\times10^{-9}$ at $N_e\sim 60$ e-folds with $\lambda/\xi^2=10^{-10}$. 

The concept of generating PBHs is examined within a class of 
single-field models of inflation. In this scenario, there is a notable contrast between 
the dynamics on small cosmological scales and the dynamics on large scales observed 
through the cosmic microwave background (CMB). This disparity proves advantageous 
in establishing the appropriate conditions for generating PBHs. 
Consequently, as the perturbed scales re-enter our Universe's horizon during later stages 
of radiation and subsequent matter dominance, these initial seeds undergo collapse, 
leading to the formation of PBHs. 

The generation of substantial scalar fluctuations during the inflationary period can lead to 
formation of significant density fluctuations, which play a vital role in the emergence 
of PBHs. The study of PBHs has garnered considerable attention over 
years, as PHBs have the potential to contribute a significant portion or even 
the entirety of the dark matter content of the Universe. 
However, the abundance of PBHs is subject to a number of stringent constraints imposed 
by their gravitational effects and evaporation rate. To produce PBHs in the early Universe, 
the magnitude of the curvature power spectrum needs to be approximately at 
the order of $10^{-3}$ to $10^{-2}$. In order to satisfy a successful inflation model, 
the curvature power spectrum is expected to yield a value of approximately 
$2.1 \times 10^{-9}$ at the scale of the CMB. 
Based on the information presented in Fig.~\ref{fig1}, it is evident that the 
basic Higgs inflation model~\cite{Bezrukov:2007ep} lacks the ability to simultaneously 
address both the inflationary period and the production of PBHs. 
Several attempts have been made to address both scenarios, 
inflation and the production of PBHs, within the framework of Higgs Inflation. 
These attempts involve the introduction of new interactions to the Higgs field. 
By incorporating these additional interactions, it is possible to achieve a successful 
inflationary period while also generating the necessary conditions for production of 
PBHs~\cite{Kawaguchi:2022nku,Ezquiaga:2017fvi,Gundhi:2020zvb,Cheong:2019vzl,Cheong:2022gfc}. 

\section{Modified Higgs Potential}\label{sec3}
This study examines a modified version of the Higgs potential that aims to address both the 
inflation and production of PBHs. Additionally, we investigate the implications of this 
modified potential in light of the recent NANOGrav signal~\cite{NANOGrav:2023ctt}. 
Here we are adding a Gaussian dip (bump)~\cite{Mishra:2019pzq} to the Higgs potential 
in Eq.~(\ref{Eq.1}). 
The structure of the Gaussian bump (dip) is given as follows
\begin{equation}\label{Eq.10a}
\pm \Bigg[A e^{-\frac{(h(\phi)-h_0(\phi))^2}{2 \sigma ^2}}\Bigg]
\end{equation}
After the conformal transformation, the potential transforms as
\begin{equation}\label{Eq.11}
    U_{eff}(\phi)=\frac{\lambda h^4(\phi)}{4}  \frac{\left(1\pm A e^{-\frac{(h(\phi)-h_0(\phi))^2}{2 \sigma ^2}}\right)}{\left(\frac{h^2(\phi) \xi }{M_{PL}^2}+1\right)^2}.
\end{equation}
The Gaussian bump (dip)  described in the above potential is featured by its height (depth) $A$ 
and position $h_0$ and width $\sigma$. The potential can be expressed in terms of the redefined field $\phi$ using equations~\ref{Eq.11} and~\ref{Eq.2}. The slow roll parameters and the power spectrum for the modified Higgs potential are given in Appendix~\ref{App.a}

\subsection{Parameter space search}
In order to investigate the appropriate parameter space of the power spectra $\mathcal{P}_{\mathcal{R}}[A,\sigma,h_0,\lambda,\xi]$ that yields a value of $\mathcal{P}_{\mathcal{R}}=10^{-2}-10^{-3}$ for 
primordial black hole (PBH) formation and $\mathcal{P}_{\mathcal{R}}=2.1\times 10^{-9}$ for a successful 
inflation model at the cosmic microwave background (CMB) scale, we conduct a scan over the 
characteristics of the bump and dip. For this analysis, we fix the Higgs self-coupling 
constant at $\lambda=0.1$ and the Higgs gravity coupling at $\xi=10^4$.

\begin{figure}[th!]
	\centering
\includegraphics[width=16.5cm,height=16cm]{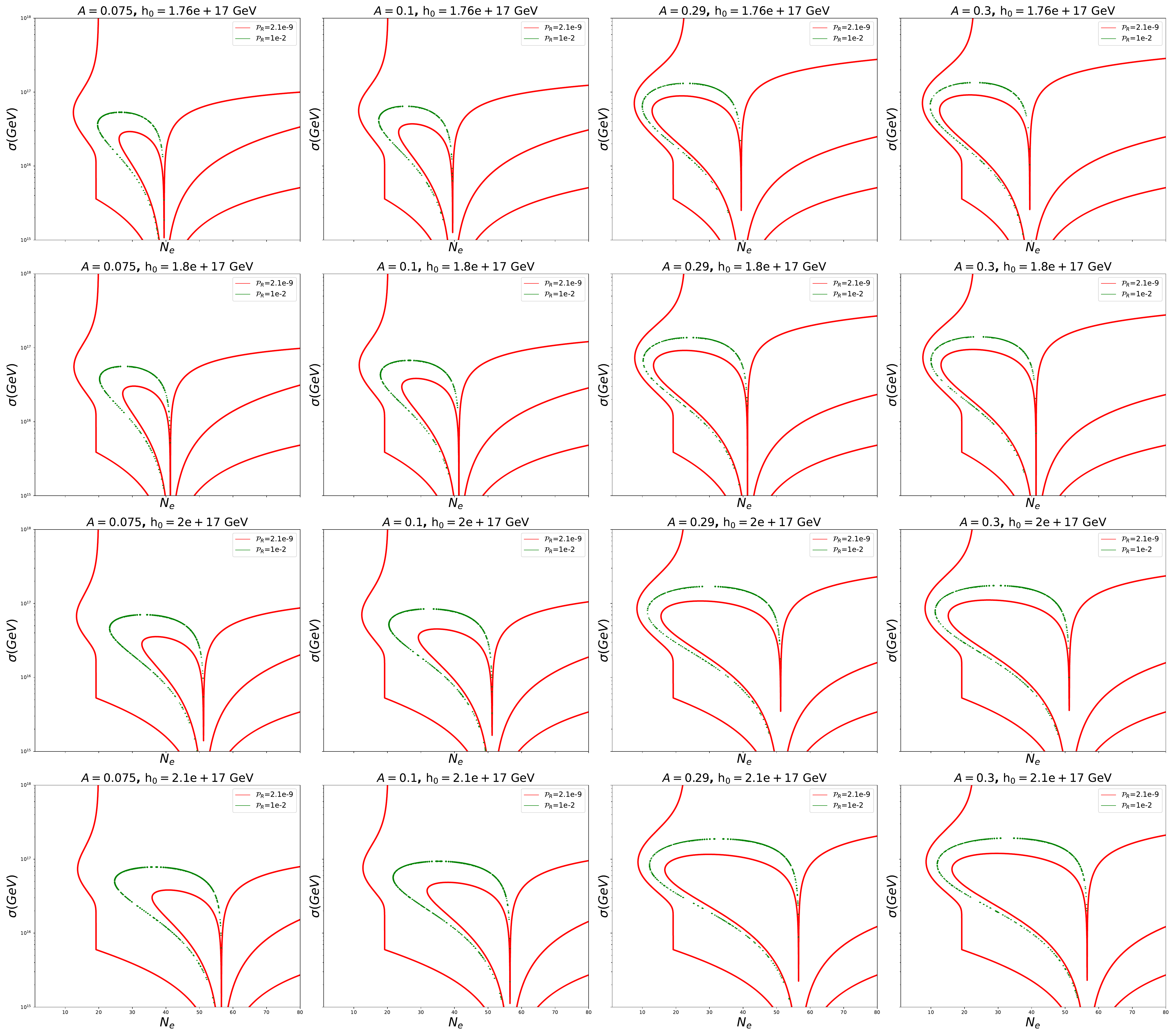}
	\caption{ \small \label{fig2} 
 The contour plot illustrates the permitted parameter space of $\sigma$ and $N_e$ for different choices of $A$ and $h_0$ 
 in the scenario of adding a dip structure,
 where the green contours correspond to $\mathcal{P}_{\mathcal{R}}=1\times 10^{-2}$ and the red contours correspond to $\mathcal{P}_{\mathcal{R}}=2.1\times 10^{-9}$.
	}
\end{figure}

Our parameter scans reveal that the addition of a dip feature to the potential 
produces a desired power spectrum for PBH formation in the early stages of inflation, 
as depicted in Fig.~\ref{fig2}. On the other hand, the inclusion of a bump feature 
in the potential only results in the required power spectrum for PBH formation at 
late CMB scales (see Appendix \ref{App.1}- Fig.~\ref{fig6}).

Figure~\ref{fig2} illustrates the parameter space of $\sigma$ and $N_e$ with varying 
values of $A$ and $h_0$. The red contour represents combinations of $\sigma$ and $N_e$ that 
yield $\mathcal{P}_{\mathcal{R}}=2.1\times 10^{-9}$ with fixed $A$ and $h_0$. 
Meanwhile, the green contour represents $\mathcal{P}_{\mathcal{R}}=1\times 10^{-2}$. We consider a 
range of $\sigma$ from $10^{15}$ to $10^{18}$ GeV and choose four arbitrary values for the depth 
of the dip $A$ (0.075, 0.1, 0.29, and 0.3). Similarly, we select four arbitrary values for 
the position of the dip $h_0$ ($1.76\times10^{17}$ GeV, $1.8\times10^{17}$ GeV, $2\times10^{17}$ GeV, and 
$2.1\times10^{17}$ GeV). For each combination we identify the corresponding values of $\sigma$ 
that yield the desired $\mathcal{P}_{\mathcal{R}}$ values for both inflation and PBH formation.

\begin{figure}[t!]
	\centering
	\includegraphics[width=15cm,height=10cm]{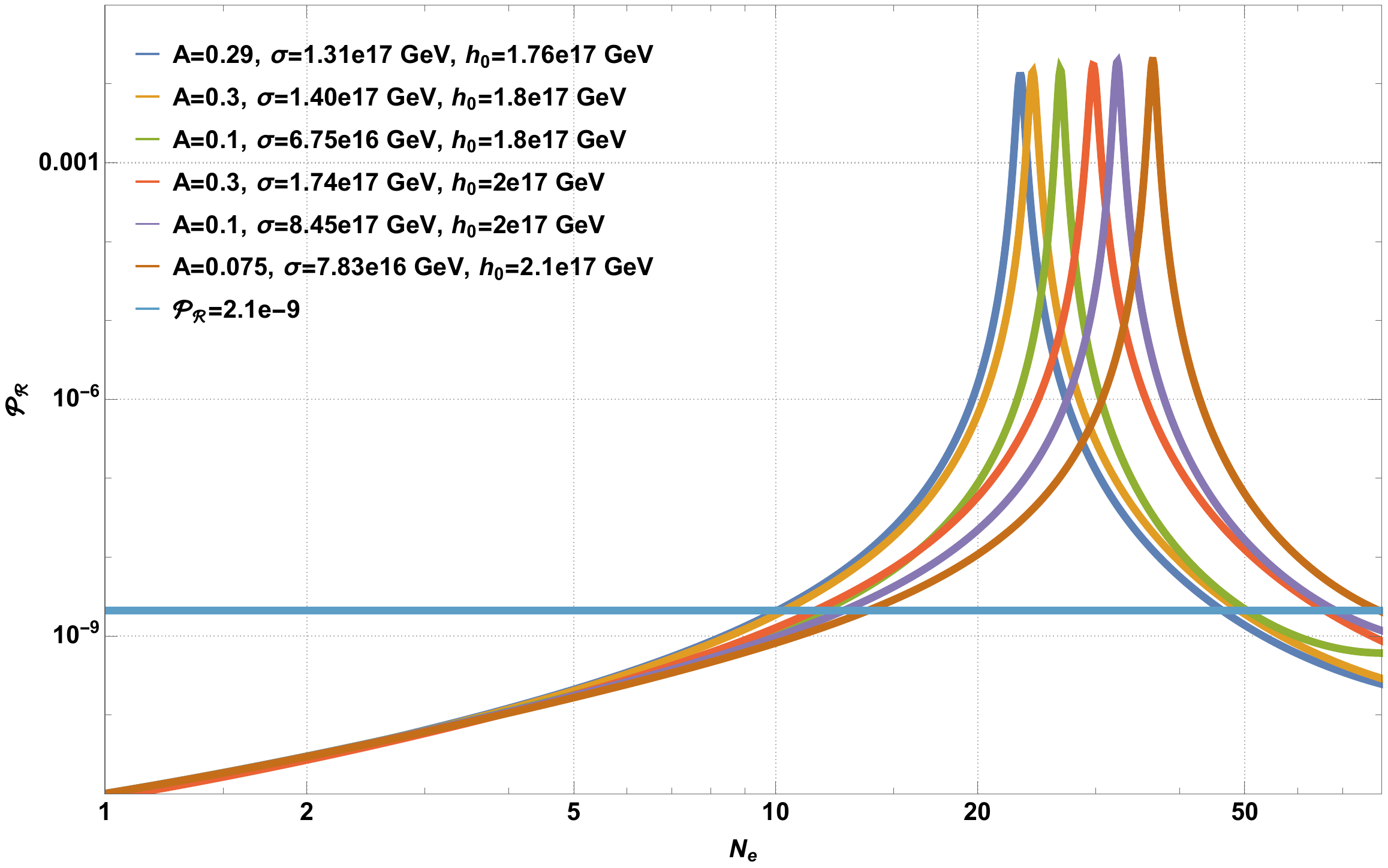}
	\caption{ \small 
 The power spectra $\mathcal{P}_{\mathcal{R}}$ versus the number of e-folds $N_e$.
 Here the power spectra exhibit new characteristics as a result of the addition of a 
 Gaussian dip in the Higgs potential. \label{fig3}
	}
\end{figure}

By observing Fig.~\ref{fig3} it becomes evident that certain parameter combinations can generate the correct power spectrum for both PBH formation and inflation at CMB scales. A comparison between Fig.~\ref{fig3} 
and Fig.~\ref{fig1} allows us to readily identify that the inclusion of a Gaussian dip in the potential has a significant impact on the power spectrum and enabling the formation of PHB seeds.

\section{PBH Formation}\label{sec4}
The PBH formation requires the power spectrum to be at least $\mathcal{O}(0.01)$, the power spectrum recorded in Fig.~\ref{fig3} guarantees this requirement. The power spectrum has a narrow peak without oscillations. The curvature perturbation $\mathcal{R}_k$ is related to the density contrast by
\begin{equation}\label{Eq.17}
\delta(t,k)=\frac{2(1+\omega)}{5+3\omega}\frac{k}{aH}\mathcal{R}_k \;.
\end{equation}
When the perturbations reenter the horizon in the radiation-dominated era, the over-dense region 
in the Universe (with $\delta>\delta_c$) would collapse into PBHs due to the increased 
amplification of curvature perturbations. This collapse occurs with $\omega=1/3$ and 
$\delta(t,k)=\frac{4}{9}\mathcal{R}_k$. Since the specifics of the PBH formation process are still 
unclear \cite{Kehagias:2019eil,Musco:2020jjb}, the precise value of the threshold $\delta_c$ remains 
uncertain. However, if we assume a Gaussian probability distribution function for the perturbations, 
the mass fraction $\beta$ of PBHs at the time of their formation can be calculated.
\begin{table}[t!]
\caption{\label{Tb.1} Inflationary observables for the different choices of the potential parameters. 
{
The resulting PBH fraction $P_{\rm PBH}$ of each region is indicated in Fig.~\ref{fig4}. 
Here $N$ is the number of e-folds, $A$, $h_0$, and $\sigma$ are the depth,
the position, and the width of the dip, respectively.
}
}
\begin{ruledtabular}
\begin{tabular}{ccccccc}
    Region & A & $\rm h_0(GeV)$ & $\rm \sigma(GeV)$ & $\rm N$ & $n_s$ & $r$ \\
    \hline
    a & 0.29 & $1.76\times10^{17}$ & $1.31\times10^{17}$ & N=47 & 0.950907 & 0.0242479 \\
    \hline
    b & 0.3  & $1.8\times10^{17}$ & $1.40\times10^{17}$ & N=50 & 0.980819 & 0.0244824 \\
    \hline
    c & 0.1  & $1.8\times10^{17}$ & $6.75\times10^{16}$ & N=51 & 0.988732 & 0.0300489 \\
    \hline
    d & 0.3  & $2\times10^{17}$ & $1.74\times10^{17}$ & N=65 & 0.98924 & 0.0206994 \\
    \hline
   e & 0.1  & $2\times10^{17}$ & $8.45\times10^{16}$ & N=68 & 0.98953  & 0.0279209 \\
    \hline
   f & 0.075 & $2.1\times10^{17}$ & $7.83\times10^{16}$ & N=78 & 0.989654 & 0.0275894\\
\end{tabular}
\end{ruledtabular}
\end{table}

\begin{equation}\label{Eq.18}
\beta(M_{PBH})=2\gamma\int_{\delta_c}^\infty \frac{d\delta}{\sqrt{2\pi}\sigma_{MPH}}exp\Big(-\frac{\delta^2}{2\sigma^2_{PBH}}\Big)\simeq\sqrt{\frac{2}{\pi}}\frac{\gamma}{\nu_c}exp\Big(-\frac{\nu^2_c}{2}\Big) \;,
\end{equation}
where $\gamma$ is the fraction of mass transformed to be PBHs that has $\delta>\delta_c$, and 
in this study, we choose $\gamma=0.4$~\cite{Gangopadhyay:2021kmf,Garcia-Bellido:2017mdw,
Garcia-Bellido:1996mdl,Kawaguchi:2022nku}. Here $\nu_c=\delta_c/\sigma_{M_{PBH}}$ and 
the variance $\sigma_{M_{PBH}}$ is defined as
\begin{equation}\label{Eq.19}
\sigma^2_{M_{PBH}}=\int_0^\infty \frac{dk}{k}\frac{16}{81}(kR)^4 W^2(kR)\mathcal{P}_{R}(k) \;,
\end{equation}
where $W(kR)$ is the window used to smooth the density contrast on comoving scale $R$. We use the Gaussian-type window function in this work,
\begin{equation}
    \label{eq.20}
    W(kR)=\exp\Big(-\frac{k^2R^2}{2}\Big).
\end{equation}
The mass fraction is ultimately determined or obtained by performing the necessary calculations or calculations based on the assumptions and considerations mentioned earlier\cite{Young:2014ana,Gu:2022pbo}.
\begin{equation}
    \label{eq.21}
    \beta(M_{PBH})=\gamma\sqrt{\frac{2}{\pi}}~\frac{4\mathcal{P}_{R}(k)}{9\delta_c}exp\Big(-\frac{81\delta^2_c}{32\mathcal{P}_{R}(k)}\Big)  \;.
\end{equation}
The mass fraction of PBHs can be related to the abundance $f_{PBH}$ as follows when considering PBHs as a fraction of dark matter:
\begin{equation}
    \label{eq.22}
  \beta(M_{PBH})=3.7\times10^{-9}\Big(\frac{\gamma}{0.2}\Big)^{-1/2}\times\Big(\frac{g_{*form}}{10.75}\Big)^{1/4}\Big(\frac{M_{PBH}}{M_{\odot}}\Big)^{1/2}f_{PBH} \;,
\end{equation}
where $M_{\odot}$ is the solar mass and $g_{*form}$ is the relativistic degrees of freedom at formation. 

The mass of PBH at the formation can be written as a fraction of horizon mass given by
\begin{equation}
    \label{eq.23}
    M_{PBH}=\gamma\frac{4\pi M^2_P}{H_N}e^{2N}
\end{equation}
where $N$ is the number of e-folds during horizon exit and $H_N$ is the Hubble expansion rate 
evaluated near the inflection point. One can calculate the mass fraction of PBHs using Eq.[~\ref{eq.21}-~\ref{eq.23}].

\begin{figure}[t!]
	\centering
	\includegraphics[width=15cm,height=10cm]{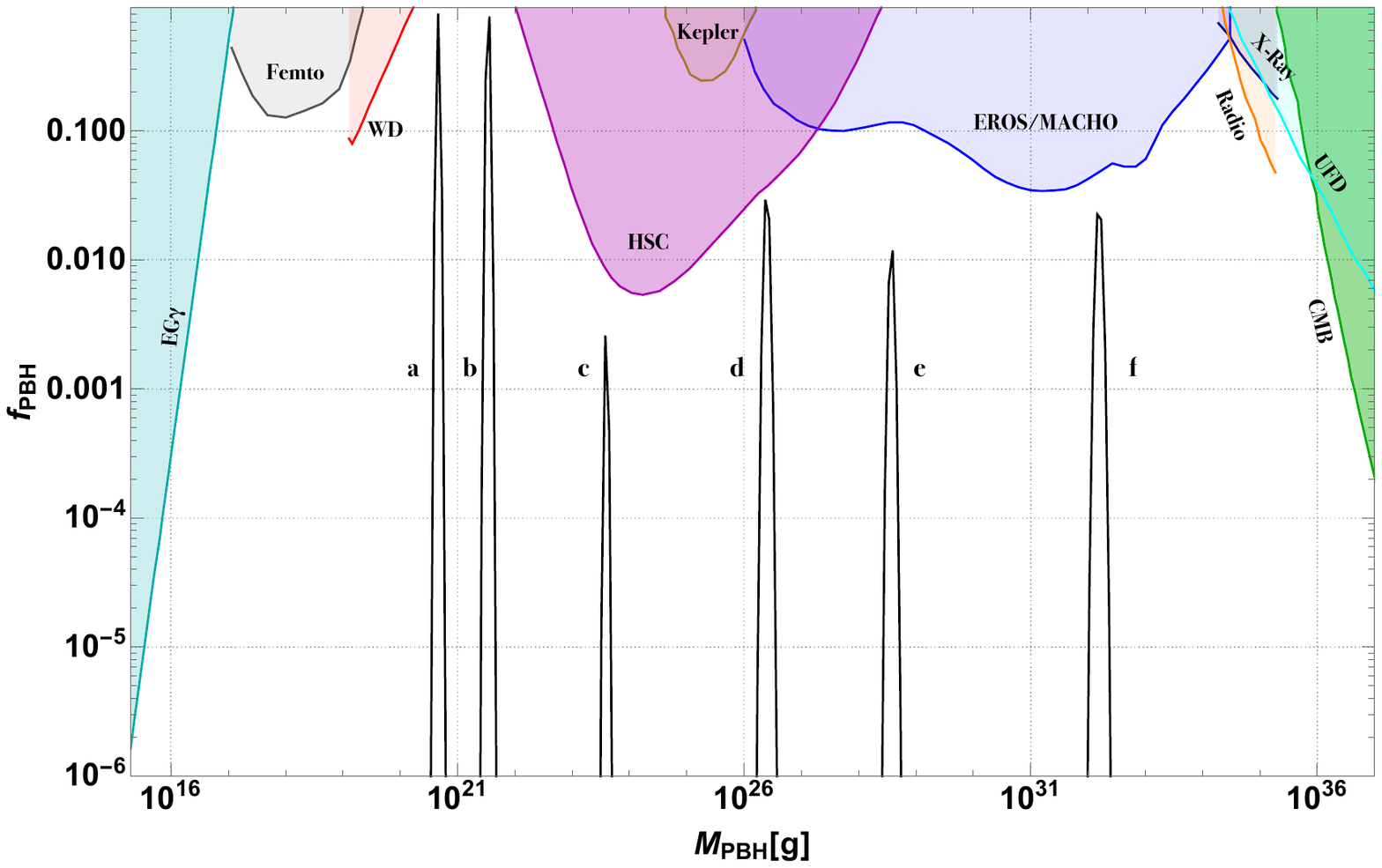}
	\caption{ \small \label{fig4} 
 The fraction of PBHs as a dark matter candidate for the parameters set 
 $\rm Region-a,b,c,d,e,$ and $f$ in Table.\ref{Tb.1}. The relevant observational constraints on the current
 primordial black hole (PBH) mass spectrum  are represented by solid lines with shades. 
 These constraints include extra-galactic gamma-ray (EG$\gamma$) observations \cite{Carr:2009jm}, 
 femtolensing data (Femto) \cite{Barnacka:2012bm}, 
 the presence of white dwarfs in our local galaxy (WD) \cite{Graham:2015apa}, 
 Subaru HSC microlensing (HSC) results \cite{Niikura:2017zjd}, 
 Kepler milli/microlensing (Kepler) measurements \cite{Griest:2013esa}, 
 EROS/MACHO microlensing observations (EROS/MACHO) \cite{EROS-2:2006ryy}, 
 dynamical heating of ultra-faint dwarf galaxies (UFD) \cite{Brandt:2016aco}, 
 constraints from X-ray/radio observations \cite{Gaggero:2016dpq}, 
 and the accretion constraints by CMB \cite{Ali-Haimoud:2016mbv,Aloni:2016kuh,Horowitz:2016lib}.}
\end{figure}

 We obtain the abundance of PBHs as dark matter, denoted as $f_{PBH}$, for a critical threshold parameter 
$\delta_c=0.414$ \cite{Gangopadhyay:2021kmf}. The corresponding results are presented in Fig.~\ref{fig4}. 
Information regarding the distinct regions marked in Fig.~\ref{fig4} can be found in Table~\ref{Tb.1}. 
A dip is 
observed with a depth of $A=0.29$ and $0.3$, accompanied by widths of $\sigma=1.31\times10^{17}$ GeV and 
$1.40\times10^{17}$ GeV, and positioned at $h_0=1.76\times10^{17}$ GeV and $1.8\times10^{17}$ GeV. 
This setting generates PBHs that potentially contribute to approximately $100\%$ of the dark matter, 
as indicated by the regions labeled $a$ and $b$ in Fig.~\ref{fig4}.

The parameters associated with regions $a$, $b$ $c$, $d$, $e$, and $f$ produce PBHs and  
spectral index $n_s$ situated on the edge of the allowed values obtained from 
cosmic microwave background (CMB) observations. Moreover, these parameters have the capability to generate 
heavier PBHs. In our study, we conducted several parameter space scans to generate PBHs. 
One common feature that emerged was an increase in the depth $A$ of the potential, 
resulting in the production of lighter PBHs while keeping other parameters fixed. 
For instance, in Table~\ref{Tb.1}, we observe this behavior in the case of region $b$ and $c$,
where $h_0$ remains fixed and $A$ ranges from 0.1 to 0.3. Similarly, decreasing the value of 
$h_0$ while keeping other parameters fixed leads to a transition that yields lighter PBHs. 
This is exemplified by the behavior of region $d$ and $b$ in Table~\ref{Tb.1}, 
where $A$ is fixed and $h_0$ varies from $2\times10^{17}$ GeV to $1.8\times10^{17}$ GeV.

Another noteworthy feature of the dip is that fixing both the depth $A$ and position $h_0$ to 
specific values while increasing the width $\sigma$ of the dip, would result in a reduction and 
eventual disappearance of the $f_{PBH}$ curve. Higher values of $\sigma$ indicate the absence of a dip, 
with the potential reverting to its original form and no dip effect. We have demonstrated this behavior 
in Appendix~\ref{App.2}, where we fixed $A=0.3$ and $h_0=1.8\times10^{17}$ GeV and progressively 
increased the width $\sigma$.

\section{Stochastic second-order gravitational wave background}\label{sec5}

Production of gravitational waves through the second-order effect occurs simultaneously with the 
formation of PBHs, specifically when the modes re-enter the Hubble radius. 
Following their production, the gravitational waves propagate freely during subsequent epochs of the 
Universe due to their low interaction rates. The frequency of these gravitational waves 
corresponds to the Hubble mass at that particular time. Considering that the mass of PBHs is 
proportional to the Hubble mass, we can establish a relationship between the PBH mass and the 
present-day frequency of gravitational waves~\cite{Sasaki:2018dmp}.

\begin{equation}
    \label{Eq.24}
    f_{GW}\simeq10^{-9}\Big(\frac{M_{PBH}}{30M_{\odot}}\Big)^{-\frac{1}{2}}~\text{Hz}
\end{equation}

\begin{figure}[th!]
	\centering
	\includegraphics[width=15cm,height=10cm]{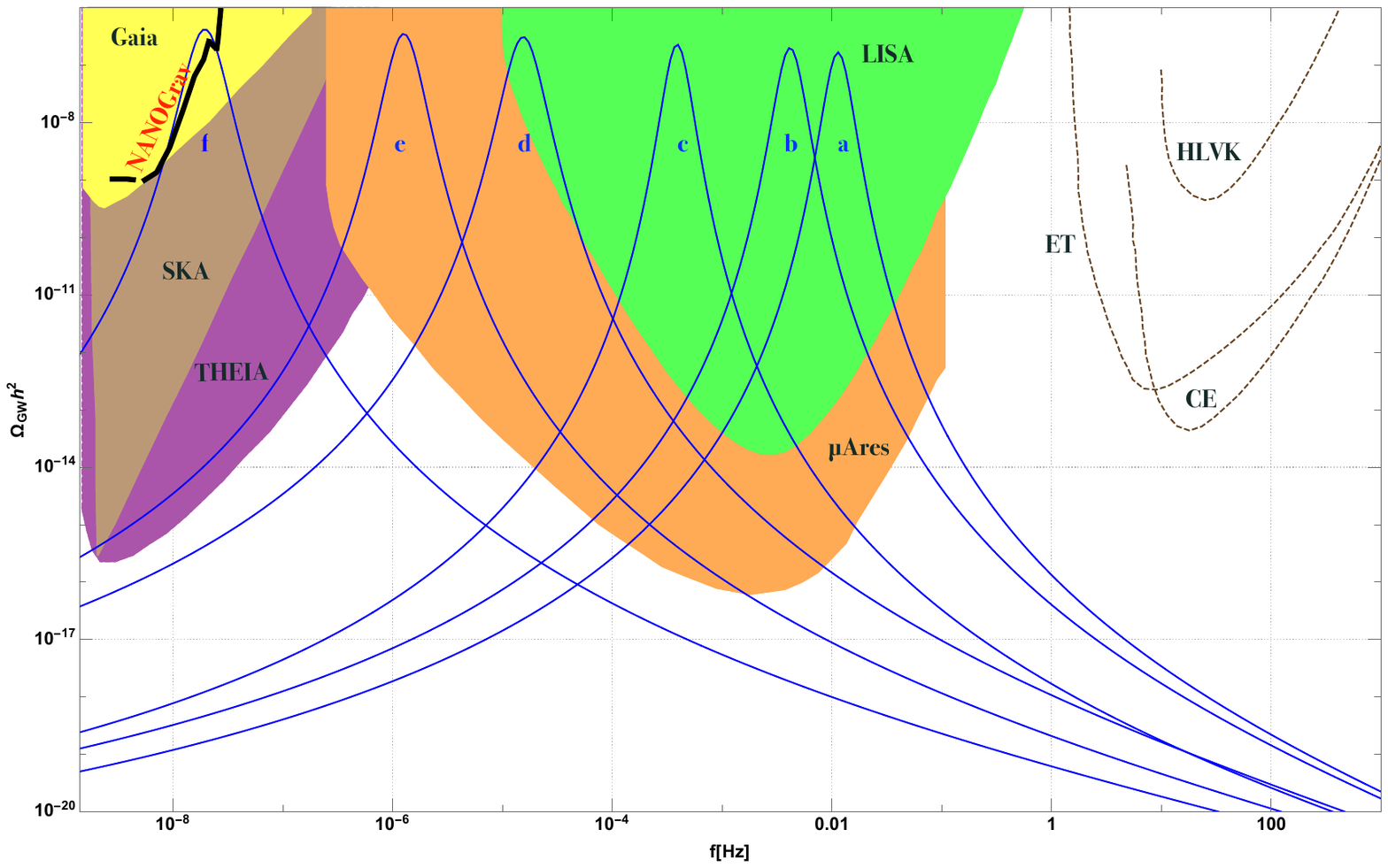}
	\caption{ \small \label{fig5} 
The gravitational wave abundance $\Omega_{\rm GW} h^2$ 
versus the frequency $f$, 
 corresponding to the benchmark parameter sets listed in Table~\ref{Tb.1}. 
 They are compared with the recent NANOGrav 15 years sensitivity~\cite{NANOGrav:2023ctt} (black curve) and
 projecting SKA/THEIA~\cite{Janssen:2014dka,Theia:2017xtk}, which utilize the observations of 
 pulsar timing array for stochastic GW of $\mathcal{O}({\rm nHz})$. 
 The planned GW interferometers
 LISA/$\mu$Ares~\cite{Caprini:2015zlo,Auclair:2019wcv,Sesana:2019vho} 
 will cover the range from $\mu$Hz to Hz.
	}
\end{figure}

The large density perturbations not only produce the PBH dark matter but also generate the second-order gravitational wave signal \cite{Matarrese:1997ay,Mollerach:2003nq}. The current relative energy density of gravitational wave obtained from the power spectra recorded in \cite{Baumann:2007zm,Di:2017ndc,Halkoaho}
\begin{equation}\label{Eq.25}
    \Omega_{GW}=10\mathcal{P}^2_{\mathcal{R}}a_{eq}
\end{equation}
We choose the current scale factor $a=1$ and $a_{eq}$ is the value of scale factor at the matter radiation 
equality defined as
\begin{equation}\label{Eq.26}
    a_{eq}=\frac{a_0}{3.1\times10^4~\Omega_Mh^2},
\end{equation}
where $h=\frac{H_0}{100km/s/Mpc}$, $H_0=67.27\,{\rm km}/s$ and $\Omega_M=0.3$. 
We plot $\Omega_{GW}h^2$ in Fig.~\ref{fig5}. 

In the main plot (Fig.~\ref{fig5}), our results indicate the values of $\Omega_{GW}h^2$, 
which are divided into several distinct regions. Each region is denoted and explained in 
Table~\ref{Tb.1}.

Region f, depicted in Fig.~\ref{fig5}, is particularly relevant as it can potentially explain the recent findings 
from NANOGrav \cite{NANOGrav:2023ctt}. The parameter space associated with region f is characterized by 
a dip in the Higgs potential with a depth of $A=0.075$ and a width of $7.83\times10^{16}$ GeV. 
This dip is positioned at $h_0=2.1\times10^{17}$ GeV. 

The relationship expressed in Eq.~(\ref{Eq.24}) reveals that $f_{GW}$ is proportional to 
$(M_{PBH})^{-1/2}$, indicating an inverse dependence between the frequency of gravitational waves 
($f_{GW}$) and the mass of primordial black holes ($M_{PBH}$). 

Additionally, it is worth noting that increasing the depth $A$ and shifting the position $h_0$ of the dip 
in the Higgs potential can lead to a shift in the corresponding curves towards higher frequency regimes. 
This observation suggests that adjustments in the parameters controlling the dip can influence the frequency
spectrum of the generated gravitational waves.
\pagebreak
\section{Conclusions}
In this study, we have investigated possible modifications to the classical Higgs inflation model proposed 
by M. Shaposhnikov and F. L. Bezrukov. We have introduced a modification to the Higgs potential by 
incorporating a dip at the top base of the potential. This modification has a significant impact on 
the generation of curvature perturbations, resulting in the amplification of second-order stochastic
gravitational wave production and potential formation of PBHs. 
These effects were absent in the classical model.

The introduction of the dip in the potential leads to an enhancement in the power spectrum, which in turn allows
for the potential existence of PBHs with masses ranging from $1.5\times10^{20}$ g to $9.72\times10^{32}$ g,
depending on the chosen parameter space. Additionally, we have demonstrated that the selected parameter values
for PBH production align with the allowed values for inflationary parameters. This suggests a consistent and
viable framework for understanding the origins of PBHs and their role as potential contributors to the dark 
matter content of our universe.

Furthermore, we have identified specific regions within the parameter space that could account 
for a significant portion of the observed dark matter. By considering various cosmological constraints,
we have established the consistency of these parameter regions. Additionally, we found that the resulting
gravitational wave signals from our model can explain the observed excess observed by NANOGrav.

In conclusion, our study highlights the importance of the modified Higgs potential in the context of the
classical Higgs inflation model. The introduced dip in the potential not only enhances the power spectrum and
allows for the formation of PBHs, but also provides a promising avenue for explaining the observed excess in gravitational wave signals and addressing the dark matter puzzle in our Universe.

\section*{Acknowledgement}
Special thanks are extended to Yogesh for engaging in an enlightening and productive discussion. 
K.C. also thanks Hyun Min Lee for the discussion related to the NANOGrav data. K.C. and C.J.O. are supported by MoST under 
Grant no. 110-2112-M-007-017-MY3. P.Y.Tseng is supported in part by the National Science and Technology Council with
Grant No. NSTC-111-2112-M-007-012-MY3. 
\vspace{-\baselineskip} 
\appendix
\section{Slow Roll Parameters and Power Spectrum for the Modified Higgs Potential}\label{App.a}
The slow-roll and other inflationary parameters with this effective potential (Eq.~\ref{Eq.11}) can be expressed as
by following Eq.~(\ref{Eq.5}) -- Eq.~(\ref{Eq.10}), 

\begin{equation}\label{Eq.13}
\epsilon =\frac{\left(A \text{Mpl}^2 \left(h \left(h^2-4 \sigma ^2\right)\pm h^2 h_0\right)+A h^4 \xi \left(h\pm h_0\right)+4 h \text{Mpl}^2 \sigma ^2 e^{\frac{\left(h- h_0\right)^2}{2 \sigma ^2}}\right)^2}{12 h^6 \xi ^2 \sigma ^4 \left(A\pm e^{\frac{\left(h-h_0\right)^2}{2 \sigma ^2}}\right)^2}
\end{equation}
Using $\epsilon$=1, we can obtain the Higgs field value at the end of inflation.
\begin{equation}\label{Eq.14}
\begin{aligned}
\eta &= \frac{\pm A \left(h^6 \xi \left(2 \left(h-2 h_0\right) \text{Mpl}^2+\xi \left(-2 h \sigma ^2+h h_0^2+h_0 \sigma ^2\right)\right)\right.}{6 h^4 h \xi ^2 \sigma ^4 \left(e^{\frac{\left(h-h_0\right)^2}{2 \sigma ^2}}\pm A\right)} \\
&\quad + \frac{\left. h^4 \text{Mpl}^2 \left(\left(h-2 h_0\right) \text{Mpl}^2+2 \xi \left(-5 h \sigma ^2+h h_0^2+4 h_0 \sigma ^2\right)\right)\right.}{6 h^4 h \xi ^2 \sigma ^4 \left(e^{\frac{\left(h-h_0\right)^2}{2 \sigma ^2}}\pm A\right)} \\
&\quad + \frac{\left. h \text{Mpl}^2 \left(h^2 \left(h_0^2 \text{Mpl}^2-8 \sigma ^2 \left(\text{Mpl}^2+\xi \sigma ^2\right)\right)+8 \text{Mpl}^2 \sigma ^4\right)\right.}{6 h^4 h \xi ^2 \sigma ^4 \left(e^{\frac{\left(h-h_0\right)^2}{2 \sigma ^2}}\pm A\right)} \\
&\quad + \frac{\left. 7 h_0 h^2 \text{Mpl}^4 \sigma ^2 + \left(h-2 h_0\right) h^8 \xi ^2\right) - 8 h \text{Mpl}^2 \sigma ^4 e^{\frac{\left(h-h_0\right)^2}{2 \sigma ^2}} \left(h^2 \xi -\text{Mpl}^2\right)}{6 h^4 h \xi ^2 \sigma ^4 \left(e^{\frac{\left(h-h_0\right)^2}{2 \sigma ^2}}\pm A\right)}
\end{aligned}
\end{equation}
The number of e-folds for each case is calculated as
\begin{equation}\label{Eq.15}
N_e=6 \xi ^2 \int_{h_{end}}^{h_{int}} \frac{h^3}{\left(h^2 \xi +\text{Mpl}^2\right)^2 \left(\frac{A \left(h^2\pm h h_0\right)}{\sigma ^2 \left(e^{\frac{\left(h-h_0\right)^2}{2 \sigma ^2}}\pm A\right)}+\frac{4 \text{Mpl}^2}{h^2 \xi +\text{Mpl}^2}\right)} \, dh
\end{equation}
The $\pm$ signs in equations~[\ref{Eq.13}]-[\ref{Eq.15}] represents the bump and dip, respectively.
Compared to the original model of inflation, the new modification generates complex values 
for $\epsilon,\eta, ~\text{and}~ N_e$. The analytical evaluation of these quantities is harder,  
and it is not possible to solve the Eq.~(\ref{Eq.15}) analytically, so we employ a numerical 
approach to compute the inflationary observables. The calculations of these observables, 
as well as the analysis of PBHs, are performed using a Python code developed by the authors. 
The power spectra are obtained as
\begin{equation}\label{Eq.16}
   \mathcal{P}_{\mathcal{R}} = \begin{cases}
\frac{h^8 \lambda  \xi ^2 \sigma ^4 e^{-\frac{\left(h-h_0\right)^2}{2 \sigma ^2}} \left(A+e^{\frac{\left(h-h_0\right)^2}{2 \sigma ^2}}\right)^3}{8 \pi ^2 \left(h^2 \xi +\text{Mpl}^2\right)^2 \left(A \left(h h_0 \left(h^2 \xi +\text{Mpl}^2\right)-h^2 \text{Mpl}^2+h^4 (-\xi )+4 \text{Mpl}^2 \sigma ^2\right)+4 \text{Mpl}^2 \sigma ^2 e^{\frac{\left(h-h_0\right)^2}{2 \sigma ^2}}\right)^2}, & \text{Potential with bump} \\
\frac{h^8 \lambda  \xi ^2 \sigma ^4 e^{-\frac{\left(h-h_0\right)^2}{2 \sigma ^2}} \left(e^{\frac{\left(h-h_0\right)^2}{2 \sigma ^2}}-A\right)^3}{8 \pi ^2 \left(h^2 \xi +\text{Mpl}^2\right)^2 \left(A \left(-h h_0 \left(h^2 \xi +\text{Mpl}^2\right)+h^2 \text{Mpl}^2+h^4 \xi -4 \text{Mpl}^2 \sigma ^2\right)+4 \text{Mpl}^2 \sigma ^2 e^{\frac{\left(h-h_0\right)^2}{2 \sigma ^2}}\right)^2}, & \text{Potential with dip}
\end{cases} 
\end{equation}
The new power spectrum is characterized by 5 parameters $[A,\sigma,h_0,\lambda,\xi]$. By tuning these variables we can obtain the proper parameter space for the inflation, PBH production, and stochastic gravitational wave background (SGWB).

\section{Bump parameter space}\label{App.1}

The addition of a Gaussian bump to the Higgs Inflation model can indeed amplify the power spectra. However, such enhancements are only observed within a specific region characterized by a large number of e-folds. For
PBH formation to occur, the inflationary power spectrum needs to be enhanced by a factor of $10^7$ within fewer than 40 e-folds of expansion.

\begin{figure}[th!]
	\centering
	\includegraphics[width=15cm,height=12cm]{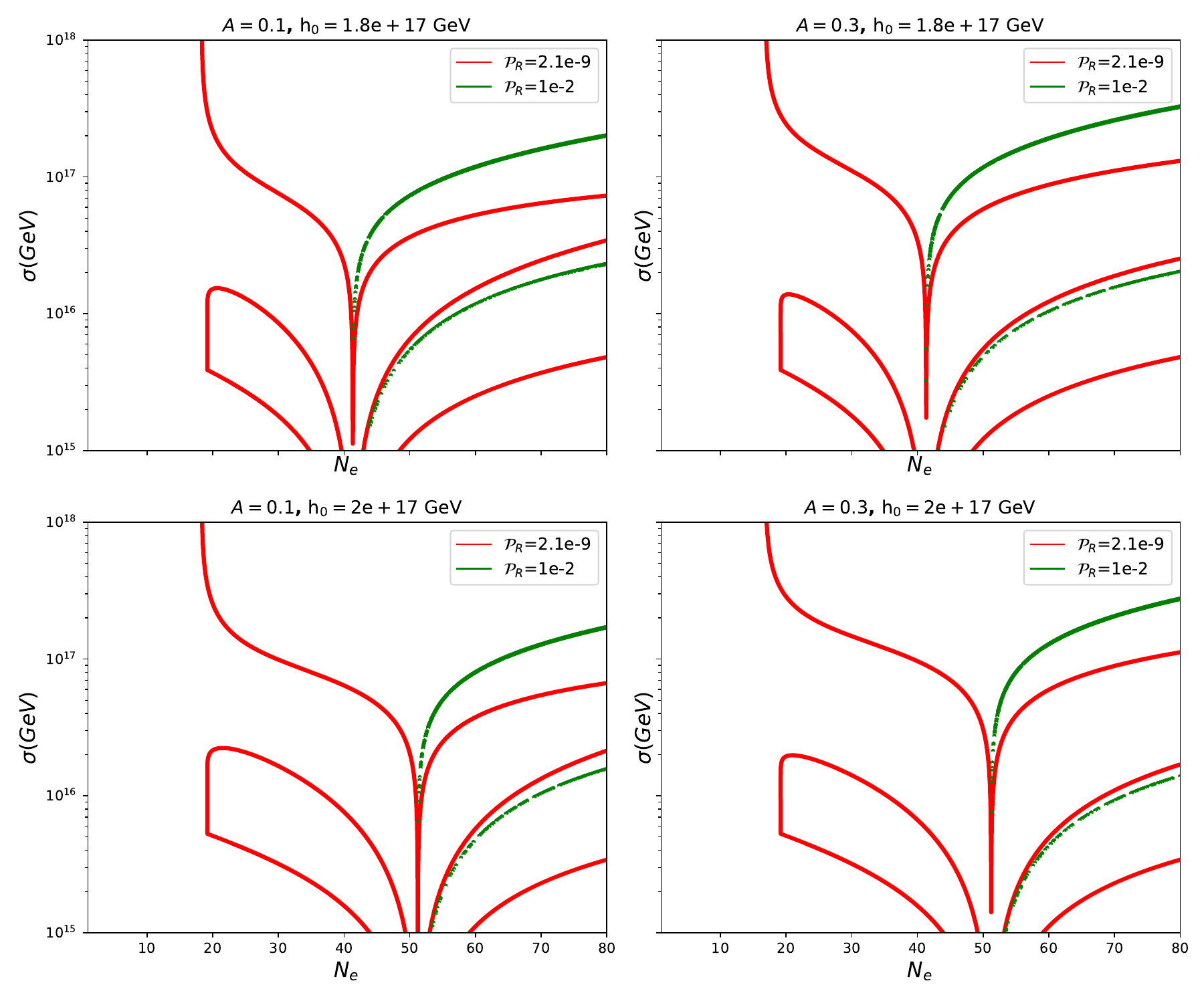}
	\caption{ \small \label{fig6} 
 The contour plot illustrates the permitted parameter space of $\sigma$ and $N_e$ for different choices of $A$ and $h_0$ 
 in the scenario of adding a bump structure,
 where the green contours correspond to $\mathcal{P}_{\mathcal{R}}=1\times 10^{-2}$ and the red contours correspond to $\mathcal{P}_{\mathcal{R}}=2.1\times 10^{-9}$
	}
\end{figure}

By comparing Figure \ref{fig2} with Figure \ref{fig6}, it becomes evident that the current bump on the potential does not provide an adequate parameter space for PBH formation. In Fig.\ref{fig6} we have shown some examples of such behavior, We explore a specific parameter space characterized by a bump with a height of $A=0.1$ (or $0.3$) and positioned at $1.8\times10^{17}$ GeV (or $2\times10^{17}$ GeV). By scanning the values of $\sigma$ and $N_e$, we generate the necessary power spectra for PBH formation. It is observed that these parameter combinations are effective for generating PBHs during later and heavier epochs of inflation, as illustrated in Figure~\ref{fig2}. Conversely, when a similar set of values is used with a dip feature in the potential, it facilitates PBH formation during smaller e-folding periods.

\section{PBH abundance Vs $\sigma$ values}\label{App.2}
\begin{figure}[t!]
	\centering
	\includegraphics[width=17cm,height=15cm]{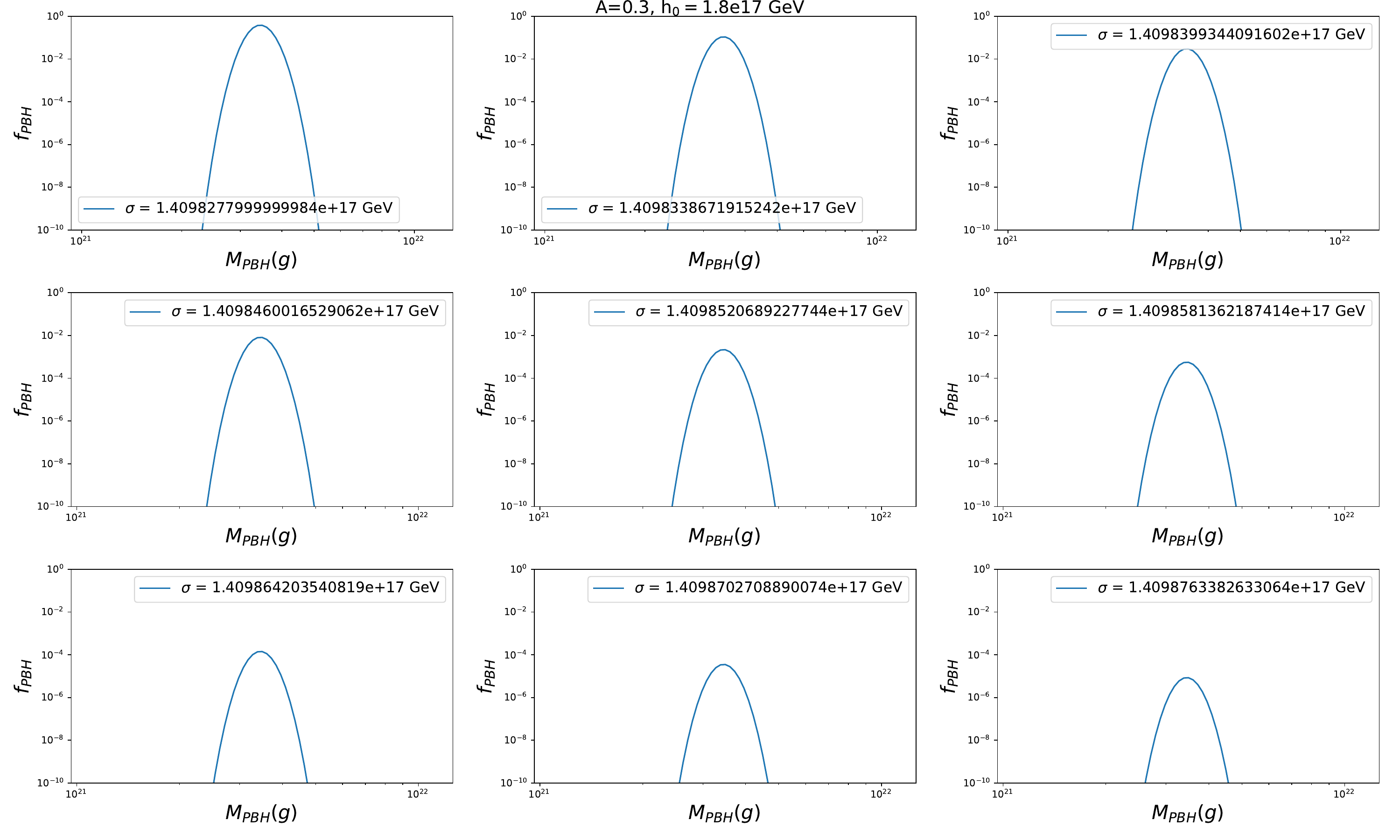}
	\caption{ \small \label{fig7} 
 The corresponding PBH abundance for the benchmark parameter $A=0.3,~h_0=1.8\times10^{17}~\rm GeV$ with the increment of the width $\sigma$ of the dip in the Higgs potential.
	}
\end{figure}

In this section, we demonstrate, using Fig.~\ref{fig7}, that when both the depth $A$ and position $h_0$ of the dip in the potential are fixed at specific values, increasing the width $\sigma$ leads to a reduction and eventual disappearance of the $f_{PBH}$ curve. Higher values of $\sigma$ indicate the absence of a dip, causing the potential to revert to its original form without a dip effect. Specifically, we set $A=0.3$ and $h_0=1.8\times10^{17}$ GeV to illustrate this feature. It is important to note that this behavior can also be observed with other choices of $A$ and $h_0$.

\end{document}